\documentclass[conference]{IEEEtran}
\IEEEoverridecommandlockouts

\usepackage{cite}
\usepackage{amsmath,amssymb,amsfonts}
\usepackage{algorithmic}
\usepackage{graphicx}
\usepackage{textcomp}
\usepackage{xcolor}
\usepackage{hyperref}
\usepackage{float}
\usepackage{multirow}

\def\BibTeX{{\rm B\kern-.05em{\sc i\kern-.025em b}\kern-.08em
    T\kern-.1667em\lower.7ex\hbox{E}\kern-.125emX}}
\begin{document}

\title{Efficient Malware Detection with Optimized Learning on High-Dimensional Features }

\author{
    \IEEEauthorblockN{Aditya Choudhary}
    \IEEEauthorblockA{\textit{Department of CSE,} \\
    \textit{School of Computational Sciences}\\
    \textit{COEP Technological University}\\
    \textit{choudharyap21.comp@coeptech.ac.in}
    }
    
    \and
    
    \IEEEauthorblockN{Sarthak Pawar}
    \IEEEauthorblockA{\textit{Department of CSE,} \\
    \textit{School of Computational Sciences}\\
    \textit{COEP Technological University}\\
    \textit{sarthaksp22.comp@coeptech.ac.in}
    }

    \and
    
    \IEEEauthorblockN{Yashodhara Haribhakta}
    \IEEEauthorblockA{\textit{Department of CSE,} \\
    \textit{School of Computational Sciences}\\
    \textit{COEP Technological University}\\
    \textit{ybl.comp@coeptech.ac.in}
    }
}

\maketitle

\begin{abstract}

Malware detection using machine learning requires feature extraction from binary files, as models cannot process raw binaries directly. A common approach involves using LIEF for raw feature extraction and the EMBER vectorizer to generate 2381-dimensional feature vectors. However, the high dimensionality of these features introduces significant computational challenges. This study addresses these challenges by applying two dimensionality reduction techniques: XGBoost-based feature selection and Principal Component Analysis (PCA). We evaluate three reduced feature dimensions (128, 256, and 384), which correspond to approximately 5.4\%, 10.8\%, and 16.1\% of the original 2381 features, across four models—XGBoost, LightGBM, Extra Trees, and Random Forest—using a unified training, validation, and testing split formed from the EMBER-2018, ERMDS, and BODMAS datasets. This approach ensures generalization and avoids dataset bias. Experimental results show that LightGBM trained on the 384-dimensional feature set after XGBoost feature selection achieves the highest accuracy of 97.52\% on the unified dataset, providing an optimal balance between computational efficiency and detection performance. The best model, trained in 61 minutes using 30 GB of RAM and 19.5 GB of disk space, generalizes effectively to completely unseen datasets, maintaining 95.31\% accuracy on TRITIUM and 93.98\% accuracy on INFERNO. These findings present a scalable, compute-efficient approach for malware detection without compromising accuracy.

\end{abstract}


\section{Introduction}

In an increasingly digital landscape, malware remains one of the most critical threats to cybersecurity. Malware—short for malicious software—exploits vulnerabilities in computer systems, often resulting in data breaches, system compromises, and other forms of cyberattacks. As both the complexity and volume of malware continue to grow, traditional detection techniques, such as signature-based methods, are proving insufficient. In response, machine learning (ML)-based malware detection has emerged as a promising alternative due to its ability to identify previously unseen or zero-day threats by learning patterns from data.

However, ML models cannot process raw malware binaries directly; these must first be converted into structured feature vectors. A widely used tool for feature extraction is LIEF \cite{b1}, which extracts static features from binaries. These features are then processed by the EMBER vectorizer to produce 2,381-dimensional feature vectors used as input for machine learning models. While this transformation makes the data suitable for training machine learning models, the high dimensionality leads to heavy computational costs, especially when dealing with large datasets.

To mitigate this, previous studies have primarily followed two approaches. The first involves training models on smaller subsets of the data to reduce computational load. However, this often limits the model’s exposure to diverse patterns, potentially causing it to miss critical features and newly emerging threats. The second approach utilizes high-end computational resources—such as GPUs or cloud infrastructure—to manage large data and feature spaces. While effective, this method is not scalable for continuous learning, where frequent updates are required to counter the evolving malware landscape. The dependency on costly infrastructure limits its practicality in many real-world scenarios.

This paper explores an alternative strategy using dimensionality reduction to enhance computational efficiency without compromising detection performance. We apply two dimensionality reduction techniques—feature selection using XGBoost and Principal Component Analysis (PCA)—to reduce the size of the original 2381-dimensional vectors. Specifically, we generate reduced versions with 128, 256, and 384 dimensions, allowing us to study how varying levels of compression affect model performance and resource usage. We then evaluate four machine learning models—XGBoost, LightGBM, Extra Trees, and Random Forest—on these reduced feature sets. Our results show that dimensionality reduction offers a practical trade-off—maintaining high detection accuracy while significantly lowering training costs—and enables efficient model updates as new malware variants emerge, making it suitable for scalable and continuous malware detection systems.

\section{Related Work}

A summary of recent malware detection studies is provided in Table \ref{tab:related_work_summary}. The EMBER-2018 dataset has been widely used to evaluate machine learning and deep learning models for static malware detection. Anderson and Roth \cite{b2} achieved strong results using LightGBM, reporting an AUC above 0.9911, outperforming early models like MalConv \cite{b3}. Wu et al. \cite{b4} improved classification accuracy from 15.75\% to 93.5\% by 

\begin{table*}[]
    \caption{Summary of Prior Malware Detection Studies}
    \label{tab:related_work_summary}
    \resizebox{1\textwidth}{!}{
        \begin{tabular}{|c|c|c|c|c|}
        \hline
        \textbf{Paper} & \textbf{Dataset} & \textbf{Approach} & \textbf{\begin{tabular}[c]{@{}c@{}}Compute\\ Environment\end{tabular}} & \textbf{\begin{tabular}[c]{@{}c@{}}Detection \\ Metric\end{tabular}} \\ \hline
        Anderson \& Roth \cite{b2} & EMBER-2018 & LightGBM & - & AUC \textgreater 0.9911 \\ \hline
        Wu et al. \cite{b4} & EMBER-2018 & \begin{tabular}[c]{@{}c@{}}Reinforcement Learning \\ with Gym-Plus\end{tabular} & - & \begin{tabular}[c]{@{}c@{}}Accuracy ↑ from \\ 15.75\% to 93.5\%\end{tabular} \\ \hline
        Oyama et al. \cite{b5} & EMBER-2018 & \begin{tabular}[c]{@{}c@{}}File Metadata + \\ Imported Functions + \\ LightGBM\end{tabular} & \begin{tabular}[c]{@{}c@{}}Intel Xeon E5-2620, \\ 128GB RAM\end{tabular} & Not specified \\ \hline
        Vinayakumar et al. \cite{b6} & EMBER-2018 & WSBD: ML + MalConv & NVIDIA India GPU grant & Accuracy: 98.9\% \\ \hline
        Pramanik \& Teja \cite{b7} & EMBER-2018 & CNN vs FFNN & - & Precision, Recall, F1: 0.97 \\ \hline
        Galen \& Steele \cite{b8} & EMBER-2018 & \begin{tabular}[c]{@{}c@{}}LightGBM on \\ time-sequenced subset\end{tabular} & - & Accuracy: 94.80\% \\ \hline
        Loi et al. \cite{b9} & EMBER-2018 & ML pipeline with static features & - & Accuracy: 96.9\% \\ \hline
        Kundu et al. \cite{b10} & EMBER-2018 & \begin{tabular}[c]{@{}c@{}}AutoML tuning\\  LightGBM hyperparameters\end{tabular} & \begin{tabular}[c]{@{}c@{}}AWS, 96/72 cores, \\ 192GB RAM\end{tabular} & \begin{tabular}[c]{@{}c@{}}TPR ↑ from \\ 86.8\% to 90\%\end{tabular} \\ \hline
        Thosar et al. \cite{b13} & EMBER-2018 & Gradient Boosting + CNN & \begin{tabular}[c]{@{}c@{}}Acer Aspire 7, \\ Intel i5 9th Gen,\\  8GB RAM\end{tabular} & Accuracy: 96\% \\ \hline
        Lad \& Adamuthe \cite{b14} & EMBER-2018 & Deep Learning (Static) & \begin{tabular}[c]{@{}c@{}}Intel Core i5-4500 CPU, \\ 8GB RAM, \\ GeForce 940M GPU\end{tabular} & Accuracy: 94.09\% \\ \hline
        Vo et al. \cite{b15} & EMBER-2018 & PEMA (XGBoost, CatBoost, LightGBM) & \begin{tabular}[c]{@{}c@{}}2× Intel Xeon Platinum, \\ 384GB RAM, 6TB SSD\end{tabular} & Accuracy: 97.65\% \\ \hline
        Shinde et al. \cite{b16} & EMBER-2018 & \begin{tabular}[c]{@{}c@{}}Random Forest + \\ Dimensionality Reduction\end{tabular} & - & Accuracy: 90\% \\ \hline
        Dener \& Gulburun \cite{b17} & EMBER-2018, BODMAS & \begin{tabular}[c]{@{}c@{}}Supervised + \\ Unsupervised (k-means + feature selection)\end{tabular} & Google Colaboratory & \begin{tabular}[c]{@{}c@{}}Accuracy: 96.77\% (EMBER), \\ 99.74\% (BODMAS)\end{tabular} \\ \hline
        Connors \& Sarkar \cite{b18} & EMBER-2018 & Neural Network Model & - & Accuracy: 95.22\% \\ \hline
        Manikandaraja et al. \cite{b19} & TRITIUM, INFERNO & \begin{tabular}[c]{@{}c@{}}Concept Drift Framework + \\ Adversarial Samples\end{tabular} & - & Not specified \\ \hline
        Rayankula \cite{b20} & BODMAS & K-Nearest Neighbors & \begin{tabular}[c]{@{}c@{}}MacOS M1, \\ 8GB RAM\end{tabular} & \begin{tabular}[c]{@{}c@{}}Accuracy: 94.9\% (Multi), \\ 94.8\% (Binary)\end{tabular} \\ \hline
        Brown et al. \cite{b22} & SOREL-20M, EMBER-2018 & AutoML (Static and Online) & \begin{tabular}[c]{@{}c@{}}92 vCPUs, \\ 448GB RAM, \\ 8× Tesla V100 GPUs\end{tabular} & Accuracy: 95.8\% \\ \hline
        Shashank et al. \cite{b24} & EMBER-2018 & Ensemble Learning (Bagging) & \begin{tabular}[c]{@{}c@{}}Nvidia DGX Station A100, \\ 4× Tesla A100 GPUs\end{tabular} & Accuracy: 96.56\% \\ \hline
        Maryam et al. \cite{b25} & EMBER-2018 & SVM (Linear SVC) & - & \begin{tabular}[c]{@{}c@{}}Accuracy: 98.9\% (14.7K samples), \\ dropped to 92.6\% (132K samples)\end{tabular} \\ \hline
        Bhardwaj et al. \cite{b26} & BODMAS & MD-ADA: Adversarial Domain Adaptation & - & \begin{tabular}[c]{@{}c@{}}Accuracy: 99.29\%, \\ F1-score: 99.13\%\end{tabular} \\ \hline
        Buriro et al. \cite{b27} & BODMAS & \begin{tabular}[c]{@{}c@{}}Anomaly Detection + \\ Random Forest\end{tabular} & - & \begin{tabular}[c]{@{}c@{}}Accuracy: 99.73\%, \\ FPR: 0\%\end{tabular} \\ \hline
        Farfoura et al. \cite{b28} & BODMAS & \begin{tabular}[c]{@{}c@{}}Dimensionality Reduction (MBMD) +\\  Random Forest\end{tabular} & - & Accuracy: 99\% \\ \hline
        \end{tabular}
    }
\end{table*}
incorporating reinforcement learning through Gym-Plus with enhanced PE modification strategies.

Later works explored combining static features such as file metadata and imported functions with ensemble models (Oyama et al. \cite{b5}), deep learning frameworks blending CNNs and MalConv (Vinayakumar et al. \cite{b6}), and AutoML-based hyperparameter tuning on LightGBM (Kundu et al. \cite{b10}), showing improvements in true positive rates at low false positive levels.

Several studies examined the trade-offs between accuracy, training time, and hardware requirements. Thosar et al. \cite{b13} and Lad \& Adamuthe \cite{b14} reported over 94\% accuracy using modest hardware. Vo et al. \cite{b15} and Shashank et al. \cite{b24} used high compute resources to reach nearly 97\%, demonstrating the impact of increased computational capacity.

Dimensionality reduction and feature selection were also crucial for handling the high-dimensional EMBER-2018 dataset, as shown by Shinde et al. \cite{b16} and Dener \& Gulburun \cite{b17}. Newer datasets like TRITIUM and INFERNO, introduced by Manikandaraja et al. \cite{b19}, include adversarial samples to study concept drift and robustness.

On related datasets like BODMAS, highly accurate models ($ >99\% $) have been developed using domain adaptation \cite{b26}, anomaly detection \cite{b27}, and novel dimensionality reduction methods \cite{b28}. Together, these works highlight the evolving landscape of static malware detection, balancing accuracy, scalability, and robustness across datasets and hardware settings.

\section{Proposed Methodology}
\subsection{Dataset}\label{subsection:dataset}
For this study, we utilized five datasets to develop and evaluate a generalized and robust malware detection model. These datasets are EMBER-2018, BODMAS, ERMDS, INFERNO, and TRITIUM, summarized in Table~\ref{table:dataset_overview}. EMBER-2018 and BODMAS contain malware samples from 2018 and 2019–2020, respectively, providing temporal diversity. ERMDS, INFERNO, and TRITIUM consist of obfuscated and adversarial malware samples exhibiting evasive behaviors, thereby increasing the model’s exposure to real-world evasion tactics and enhancing its generalization.

\begin{table}[]
\caption{Dataset Summary with Class-wise Sample Distribution}
\resizebox{0.5\textwidth}{!}{
\begin{tabular}{|l|l|l|}
\hline
\multicolumn{1}{|c|}{\textbf{Dataset}}               & \multicolumn{1}{c|}{\textbf{Description}}                                                                                                                                                                    & \multicolumn{1}{c|}{\textbf{\begin{tabular}[c]{@{}c@{}}Data \\ Distribution\end{tabular}}} \\ \hline
\begin{tabular}[c]{@{}l@{}}EMBER-2018\end{tabular} & \begin{tabular}[c]{@{}l@{}}EMBER-2018 is a large-scale dataset \\ containing featuresfrom 1 million \\ Windows PE files scanned before \\ 2018, designed for challenging \\ malware detection tasks.\end{tabular} & \begin{tabular}[c]{@{}l@{}}400K benign\\ 400K malicious\\ 300K unlabelled\end{tabular}     \\ \hline
BODMAS                                               & \begin{tabular}[c]{@{}l@{}}BODMAS is a dataset collected \\ between August 2019 and \\ September 2020.\end{tabular}                                                                                          & \begin{tabular}[c]{@{}l@{}}77,142 benign\\ 57,293 malicious\end{tabular}                         \\ \hline
ERMDS                                               & \begin{tabular}[c]{@{}l@{}}ERMDS is a dataset created to \\ assess the effectiveness of \\ learning-based malware \\ detection systems against \\ obfuscated malware samples.\end{tabular}                   & \begin{tabular}[c]{@{}l@{}}30,455 benign\\ 86,685 malicious\end{tabular}                         \\ \hline
TRITIUM                                              & \begin{tabular}[c]{@{}l@{}}TRITIUM was introduced to\\ analyze concept drift by \\ evaluating threats from a \\ distinct source and threat profile.\end{tabular}                                             & \begin{tabular}[c]{@{}l@{}}10,785 benign\\ 12,471 malicious\end{tabular}                         \\ \hline
INFERNO                                             & \begin{tabular}[c]{@{}l@{}}INFERNO is a dataset \\ developed to assess classifier \\ robustness against adversarial \\ malware.\end{tabular}                                                                 & \begin{tabular}[c]{@{}l@{}}1430 benign\\ 1430 malicious\end{tabular}                       \\ \hline
\end{tabular}
}
\label{table:dataset_overview}
\end{table}

A unified train-validation-test split with stratified sampling was established using EMBER-2018, BODMAS, and ERMDS to encourage generalization and reduce dataset-specific bias. Table~\ref{table:dataset_splits} shows the distribution of this unified dataset. Along with the unified test split, the INFERNO and TRITIUM datasets were reserved exclusively for evaluation to test the model’s performance on unseen and adversarial samples.

\subsection{Data Preprocessing and Dimensionality Reduction}\label{subsection:data_preprocessing}
\begin{figure}[]
\centering
\includegraphics[width=0.5\textwidth]{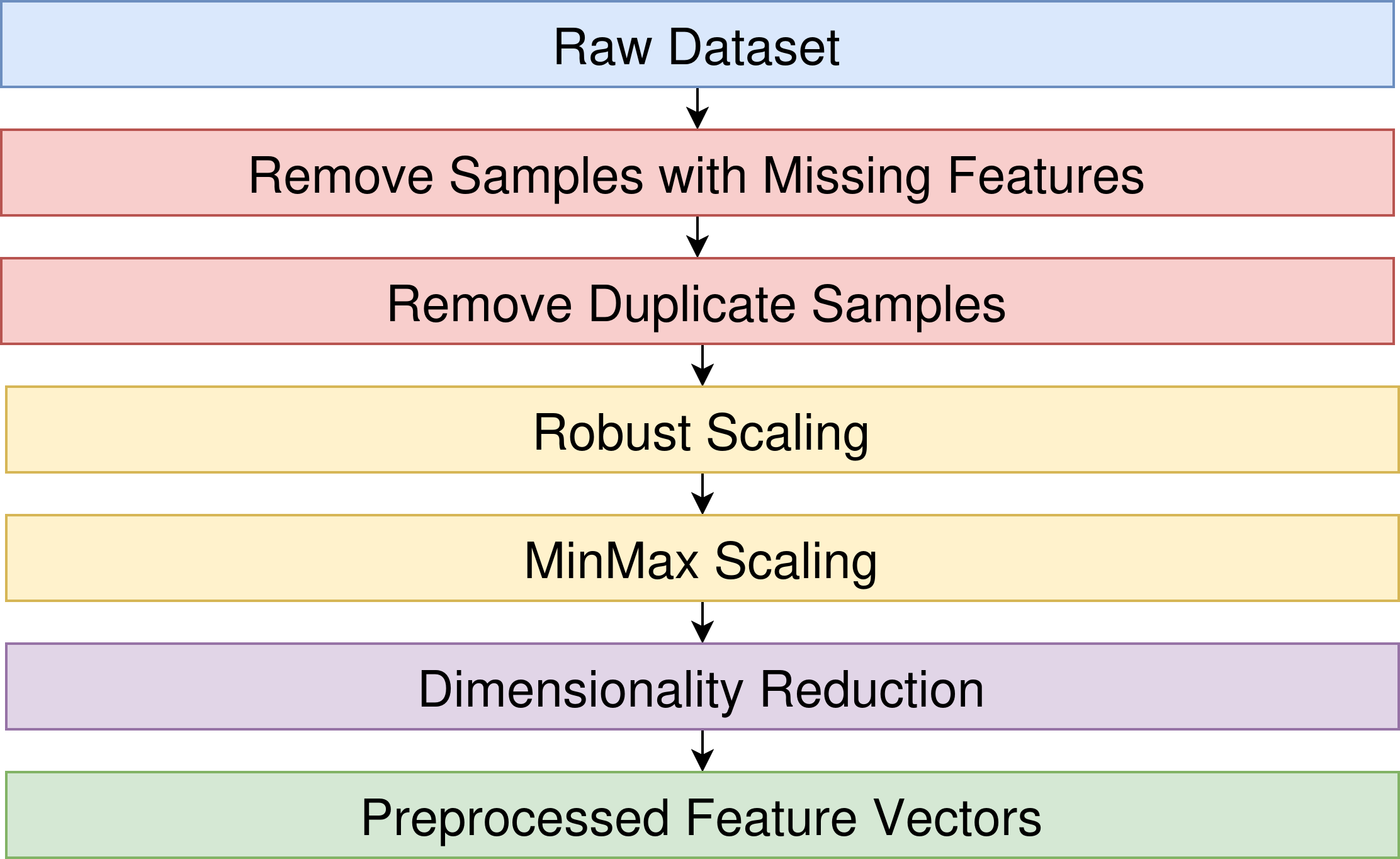}
\caption{Preprocessing Pipeline Overview}
\label{fig:preprocessing_pipeline}
\end{figure}

The datasets used in this study already contained 2,381-dimensional feature vectors extracted using the LIEF tool and EMBER vectorizer methodology described in \cite{b2}. As shown in Figure~\ref{fig:preprocessing_pipeline}, preprocessing began with the raw dataset by removing samples with missing features, followed by the elimination of duplicate entries. These steps ensured data quality and consistency before further processing.

Next, the cleaned feature vectors underwent a two-stage normalization process. First, RobustScaler was applied to reduce the influence of outliers by scaling features based on the median and interquartile range, effectively addressing the skewed distributions common in malware data. Then, MinMaxScaler scaled the features to a uniform range of [0, 1], promoting numerical stability and consistency across all feature dimensions. This combined scaling approach produced input data that was both robust to extreme values and well-normalized, facilitating efficient and stable model training.

\begin{table}[h!]
\caption{Train, Validation, and Test Splits for Unified Dataset}
\resizebox{0.5\textwidth}{!}{
    \begin{tabular}{|c|c|c|c|}
    \hline
    \textbf{Dataset}                                      & \textbf{\begin{tabular}[c]{@{}c@{}}Number of \\ Train Samples\end{tabular}} & \textbf{\begin{tabular}[c]{@{}c@{}}Number of \\ Validation Samples\end{tabular}} & \textbf{\begin{tabular}[c]{@{}c@{}}Number of \\ Test Samples\end{tabular}} \\ \hline
    \begin{tabular}[c]{@{}c@{}}EMBER-2018\end{tabular} & 479936                                                                      & 119984                                                                           & 199956                                                                     \\ \hline
    BODMAS                                                & 36662                                                                       & 9166                                                                             & 11458                                                                      \\ \hline
    ERMDS                                                 & 62748                                                                       & 15687                                                                            & 19609                                                                      \\ \hline
    \textbf{TOTAL}                                        & \textbf{579346}                                                             & \textbf{144837}                                                                  & \textbf{231023}                                                            \\ \hline
    \end{tabular}
}
\label{table:dataset_splits}
\end{table}

Finally, to improve computational efficiency without compromising classification performance, dimensionality reduction was applied using two separate techniques: feature selection via XGBoost and Principal Component Analysis (PCA). The original 2,381-dimensional feature space was independently reduced to lower dimensions of 128, 256, and 384 for each technique. For every reduced dimension and method, classification models were trained and evaluated to identify the optimal balance between dimensionality and model accuracy. This approach ensured that the reduced feature sets retained sufficient informative value while significantly lowering computational overhead during training.

\subsection{Model Training and Evaluation on Reduced Features}\label{subsection:model_training}

\begin{figure}[h]
\centering
\includegraphics[width=0.5\textwidth]{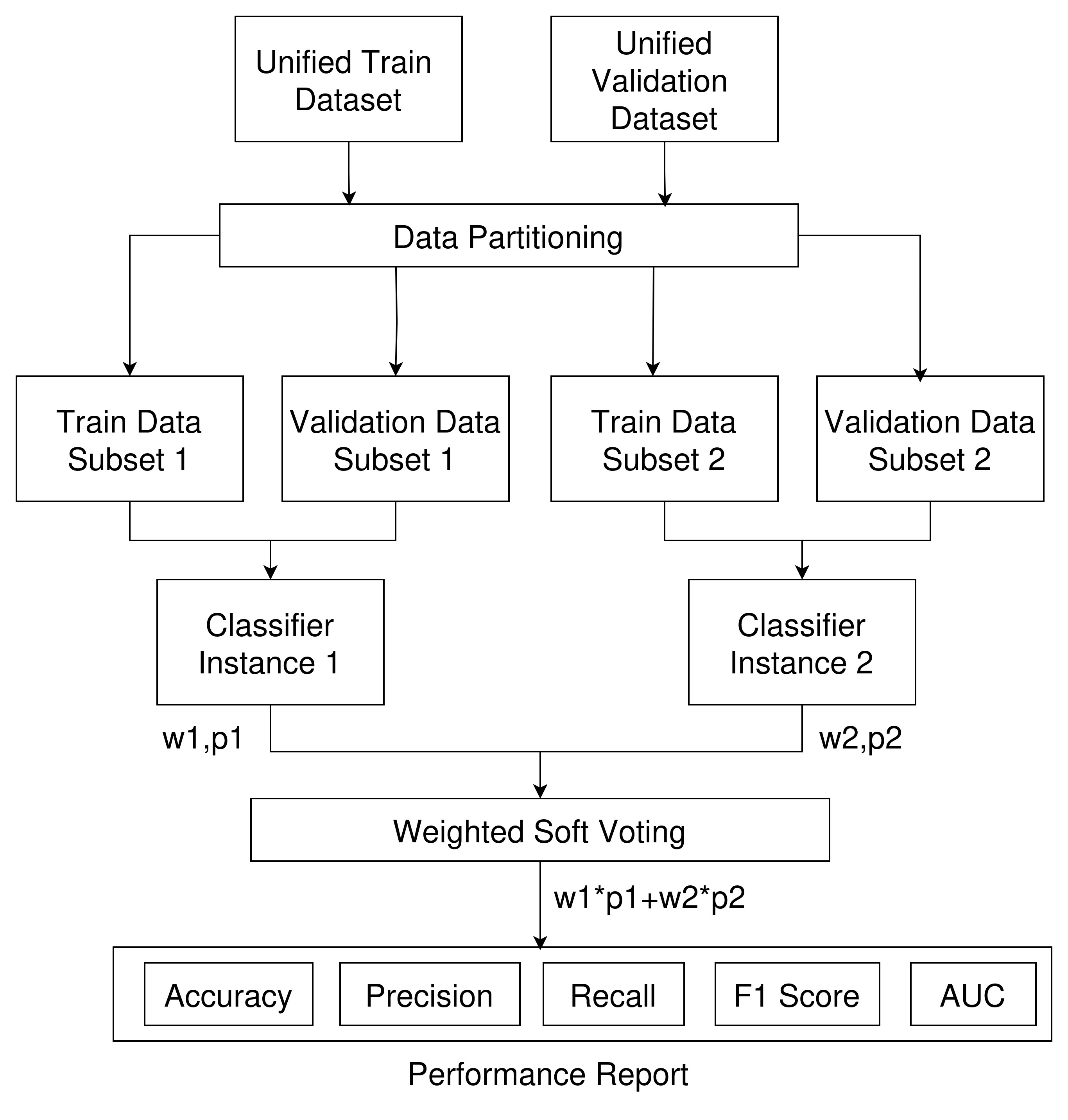}
\caption{Proposed system architecture for training and evaluating models on reduced feature sets}
\label{fig:proposed_system}
\end{figure}

After completing the preprocessing steps (Figure~\ref{fig:preprocessing_pipeline}), the processed feature vectors were used in the training and evaluation framework, as illustrated in Figure~\ref{fig:proposed_system}. To manage computational constraints, the training and validation data are divided into two equal stratified partitions.

In this study, we evaluate four machine learning classifiers—XGBoost, LightGBM, Extra Trees, and Random Forest—training each independently on both partitions, resulting in two instances per classifier. Hyperparameter tuning for each instance of model is conducted using FLAML \cite{b30} to optimize performance.

The outputs of the two instances of each model are combined using a weighted soft voting mechanism. Let \( p_1 \) and \( p_2 \) denote the probability scores from the first and second model instances, respectively. The final prediction \( \hat{p} \) is computed as:
\[
\hat{p} = w_1 \cdot p_1 + w_2 \cdot p_2
\]
where \( w_1 + w_2 = 1 \). Optimal weights are identified by iterating \( w_1 \) from 0 to 1 in increments of 0.1 during training. The weight pair \( (w_1, w_2) \) yielding the highest performance is selected and consistently applied during validation and testing. Model effectiveness is assessed using accuracy, precision, recall, F1-score, and area under the ROC curve (AUC).

\section{Results and Discussion}  
\subsection{Experimental Setup}\label{subsection:exp_setup}
All experiments were conducted using the free resources provided by Kaggle\footnote{\url{"https://www.kaggle.com}}. The computational setup included an Intel Xeon 2.20 GHz CPU, 30 GB of RAM, and 19.5 GB of disk space. Several Python libraries were employed throughout the implementation: pandas for dataset handling and operations, numpy for numerical computations, scikit-learn for preprocessing (including scalers and PCA), baseline classifier implementations, and evaluation metrics, and flaml for automated hyperparameter tuning.

\subsection{Performance on Unified Test Data}\label{subsection:perf_unified_test}

Table~\ref{table:unified_test_data_perfomance} summarizes the performance of four classifiers—XGBoost, LightGBM, Extra Trees, and Random Forest—trained on feature sets reduced to 128, 256, and 384 dimensions using two techniques: supervised feature selection with XGBoost and unsupervised PCA. 

Among the models using XGBoost-based feature selection, LightGBM demonstrated the highest performance with 384 features, achieving 97.52\% accuracy, 97.73\% F1-score, and 99.58\% AUC. XGBoost followed closely with 97.25\% accuracy and 97.48\% F1-score. Extra Trees and Random Forest also performed competitively, particularly at higher feature counts. The steady improvement from 128 to 384 dimensions indicates that additional features contributed discriminative value, enhancing classification performance. These results confirm that supervised selection effectively retains task-relevant features, benefiting tree-based classifiers.The complete training workflow for the best-performing LightGBM model with 384 features was executed in approximately 61 minutes.

In contrast, PCA-based models showed comparatively lower performance. The best PCA result—XGBoost with 384 components—reached 95.17\% accuracy and 95.55\% F1-score. Other models exhibited similar trends, with performance increasing from 128 to 384 dimensions but remaining below the XGBoost selection baseline. Random Forest was most affected, with its accuracy dropping to 91.78\%. Since PCA preserves variance without considering class labels, it may discard critical features necessary for malware discrimination, weakening downstream classifier performance.

Overall, XGBoost-based feature selection consistently outperformed PCA, confirming the importance of label-aware dimensionality reduction in malware detection. While increasing dimensions led to better results in both methods, the performance gains tended to saturate beyond 256 features, especially for PCA. This suggests a diminishing return on including more components, and underscores the benefit of selecting fewer, yet more informative, features. The combination of supervised selection and tree-based classifiers offers a robust, efficient, and scalable solution well-suited for real-time detection under limited computational resources.

\subsection{Performance on TRITIUM and INFERNO Dataset}\label{subsection:perf_unseen_data}
As shown in Table~\ref{table:unseen_data_evaluation}, we further assessed the robustness and generalization capability of our model on the TRITIUM and INFERNO datasets, which were not used during training or validation. These datasets contain obfuscated and adversarial malware samples, making them ideal benchmarks for evaluating real-world detection effectiveness.

We employed our best-performing pipeline—XGBoost-based feature selection reducing the input to 384 dimensions, followed by LightGBM classification—for this evaluation. On the TRITIUM dataset, the model achieved 95.31\% accuracy, with a precision of 99.56\%, recall of 91.75\%, and F1-score of 95.49\%. An AUC of 99.79\% further demonstrated strong separability between benign and malicious samples. On the INFERNO dataset, the model maintained robust performance, achieving 93.98\% accuracy, 91.99\% precision, 96.04\% recall, 93.97\% F1-score, and an AUC of 98.29\%.These results highlight the pipeline’s strong generalization across unseen, evasive threats and reinforce its practical applicability for real-time malware detection in dynamic and adversarial cybersecurity environments.

\begin{table*}[]
\caption{Performance comparison of models using different dimensionality reduction techniques and feature sizes on unified test data}
\resizebox{1\textwidth}{!}{
\begin{tabular}{|c|c|c|c|c|c|c|c|}
\hline
\textbf{\begin{tabular}[c]{@{}c@{}}Reduction \\ Technique\end{tabular}}                                  & \textbf{\begin{tabular}[c]{@{}c@{}}Reduced \\ Dimension\end{tabular}} & \textbf{Model} & \textbf{Accuracy}     & \textbf{Precision}    & \textbf{Recall}     & \textbf{F1 score}      & \textbf{AUC score}     \\ \hline
\multirow{12}{*}{\begin{tabular}[c]{@{}c@{}}Feature \\ Selection \\ using \\ XGBoost \end{tabular}} & \multirow{4}{*}{128}                                         & XGBoost            & 96.61\%          & 96.51\%          & 97.28\%          & 96.89\%          & 99.40\%          \\ \cline{3-8} 
                                                                                                &                                                              & LightGBM           & 96.73\%          & 96.78\%          & 97.23\%          & 97.00\%          & 99.40\%          \\ \cline{3-8} 
                                                                                                &                                                              & Extra Trees            & 96.42\%          & 96.76\%          & 96.65\%          & 96.70\%          & 99.34\%          \\ \cline{3-8} 
                                                                                                &                                                              & Random Forest             & 95.44\%          & 95.10\%          & 96.60\%          & 95.84\%          & 99.10\%          \\ \cline{2-8} 
                                                                                                & \multirow{4}{*}{256}                                         & XGBoost            & 97.05\%          & 96.96\%          & 97.64\%          & 97.30\%          & 99.51\%          \\ \cline{3-8} 
                                                                                                &                                                              & LightGBM           & 97.34\%          & 97.52\%          & 97.59\%          & 97.56\%          & 99.54\%          \\ \cline{3-8} 
                                                                                                &                                                              & Extra Trees            & 96.79\%          & 97.05\%          & 97.05\%          & 97.05\%          & 99.46\%          \\ \cline{3-8} 
                                                                                                &                                                              & Random Forest             & 95.80\%          & 95.59\%          & 96.73\%          & 96.16\%          & 99.17\%          \\ \cline{2-8} 
                                                                                                & \multirow{4}{*}{384}                                         & XGBoost            & 97.25\%          & 97.34\%          & 97.61\%          & 97.48\%          & 99.57\%          \\ \cline{3-8} 
                                                                                                &                                                              & \textbf{LightGBM}  & \textbf{97.52\%} & \textbf{97.54\%} & \textbf{97.91\%} & \textbf{97.73\%} & \textbf{99.58\%} \\ \cline{3-8} 
                                                                                                &                                                              & Extra Trees            & 97.01\%          & 97.17\%          & 97.33\%          & 97.25\%          & 99.46\%          \\ \cline{3-8} 
                                                                                                &                                                              & Random Forest             & 94.96\%          & 94.44\%          & 96.41\%          & 95.41\%          & 98.86\%          \\ \hline
\multirow{12}{*}{PCA}                                                                           & \multirow{4}{*}{128}                                         & XGBoost            & 94.85\%          & 95.38\%          & 95.13\%          & 95.25\%          & 98.89\%          \\ \cline{3-8} 
                                                                                                &                                                              & LightGBM           & 94.80\%          & 95.39\%          & 95.02\%          & 95.21\%          & 98.85\%          \\ \cline{3-8} 
                                                                                                &                                                              & Extra Trees            & 94.16\%          & 95.03\%          & 94.19\%          & 94.61\%          & 98.71\%          \\ \cline{3-8} 
                                                                                                &                                                              & Random Forest             & 92.35\%          & 92.36\%          & 93.68\%          & 93.02\%          & 97.88\%          \\ \cline{2-8} 
                                                                                                & \multirow{4}{*}{256}                                         & XGBoost            & 95.14\%          & 95.63\%          & 95.43\%          & 95.53\%          & 98.98\%          \\ \cline{3-8} 
                                                                                                &                                                              & LightGBM           & 95.00\%          & 95.56\%          & 95.23\%          & 95.39\%          & 98.92\%          \\ \cline{3-8} 
                                                                                                &                                                              & Extra Trees            & 92.69\%          & 92.99\%          & 93.61\%          & 93.30\%          & 98.05\%          \\ \cline{3-8} 
                                                                                                &                                                              & Random Forest             & 91.47\%          & 91.57\%          & 92.86\%          & 92.21\%          & 97.49\%          \\ \cline{2-8} 
                                                                                                & \multirow{4}{*}{384}                                         & \textbf{XGBoost}   & \textbf{95.17\%} & \textbf{95.63\%} & \textbf{95.48\%} & \textbf{95.55\%} & \textbf{98.95\%} \\ \cline{3-8} 
                                                                                                &                                                              & LightGBM           & 94.85\%          & 95.64\%          & 94.84\%          & 95.24\%          & 98.84\%          \\ \cline{3-8} 
                                                                                                &                                                              & Extra Trees            & 94.01\%          & 94.91\%          & 94.04\%          & 94.47\%          & 98.59\%          \\ \cline{3-8} 
                                                                                                &                                                              & Random Forest             & 91.78\%          & 91.61\%          & 93.44\%          & 92.52\%          & 97.73\%          \\ \hline
\end{tabular}
}
\label{table:unified_test_data_perfomance}
\end{table*}

\subsection{Comparison with Existing Methods}
A detailed comparison of our approach with existing methods is presented in Table~\ref{tab:comparison_with_other}. Since many prior studies benchmarked their models using the EMBER-2018 dataset, we evaluated our best-performing model (LightGBM trained on 384 features selected via XGBoost-based feature selection) on the same dataset for a fair comparison. Our model achieves a favorable balance between detection accuracy and computational efficiency.

For instance, Vo et al.~\cite{b15} attained 97.65\% accuracy using 2,291 features on high-end hardware with 384 GB RAM and Xeon Platinum CPUs. In contrast, our approach achieves a comparable 97.25\% accuracy using only 384 features and 30 GB of RAM. Similarly, Shashank et al.~\cite{b22} reported 95.80\% accuracy using 448 GB RAM and 8 Tesla V100 GPUs, whereas our method surpasses this performance without any GPU dependency. Furthermore, Dener and Gulburun~\cite{b17} achieved 96.77\% accuracy with 448 features, while our model achieves similar performance using fewer features, highlighting its efficiency.

Overall, our approach demonstrates strong predictive performance with minimal computational overhead, reduced feature complexity, and broader scalability—making it a viable alternative to more resource-intensive methods in real-world deployment scenarios.

\section{Conclusion and Future Work}
In conclusion, this study presents an optimized machine learning pipeline for static malware detection by applying dimensionality reduction techniques to high-dimensional feature vectors extracted from PE files. Specifically, XGBoost-based feature selection and Principal Component Analysis (PCA) were used to reduce the original 2,381-dimensional vectors to 128, 256, and 384 dimensions. Among the evaluated classifiers, LightGBM trained on 384 features selected via XGBoost achieved the highest performance, reaching 97.52\% accuracy and strong precision, recall, and F1-score values. The complete training workflow was executed in approximately 61 minutes, demonstrating the practicality of the proposed approach for efficient model development.

For future work, this system can be extended to support malware family classification and made more resilient through adversarial training. Incorporating deep learning-based dimensionality reduction techniques such as autoencoders and applying ensemble learning across heterogeneous classifiers may further enhance detection capabilities. Additionally, leveraging transfer learning could improve adaptability to evolving malware threats and support generalization across diverse datasets.

\begin{table}[h]
\caption{Evaluation of LightGBM with 384 features selected via XGBoost-based feature selection on TRITIUM and INFERNO datasets}
\resizebox{0.5\textwidth}{!}{
\begin{tabular}{|c|c|c|c|c|c|}
\hline
\textbf{Dataset} & \textbf{Accuracy} & \textbf{Precision} & \textbf{Recall} & \textbf{F1 Score} & \textbf{AUC Score} \\ \hline
TRITIUM          & 95.31\%           & 99.56\%            & 91.75\%         & 95.49\%           & 99.79\%            \\ \hline
INFERNO          & 93.98\%           & 91.99\%            & 96.04\%         & 93.97\%           & 98.29\%            \\ \hline
\end{tabular}
}
\label{table:unseen_data_evaluation}
\end{table}

\begin{table*}[]
\caption{Comparison of various approaches on the EMBER-2018 dataset}
\resizebox{1\textwidth}{!}{
    \begin{tabular}{|c|c|c|c|c|}
    \hline
    \textbf{Model} & \textbf{Test Data Size} & \textbf{Feature Size} & \textbf{Experimental Setup} & \textbf{Test Accuracy} \\ \hline
    WSBD model & 56,000 & 2351 & Recieved GPU Grant from NVIDIA India & 98.90\% \\ \hline
    ML pipeline with static features & 2,00,000 & 2351 & - & 96.90\% \\ \hline
    Gradient Boosting + CNN & - & 2351 & \begin{tabular}[c]{@{}c@{}}8GB RAM, 512 SSD, Nvidia GeForce GTX 1650,\\ Intel i5 9th gen processor\end{tabular} & 96.00\% \\ \hline
    Deep Learning (Static) & 2,00,000 & 2381 & \begin{tabular}[c]{@{}c@{}}Nvidia GeForce 940M 2GB GPU ,Intel Core i5-4500 processor,\\ 8 GB RAM,Google Colab\end{tabular} & 94.09\% \\ \hline
    PEMA (XGBoost, CatBoost, LightGBM) & 2,00,000 & 2291 & 2× Intel Xeon Platinum 8160 ,384GB RAM,6TB SSD & 97.65\% \\ \hline
    \multirow{2}{*}{k-means + feature selection} & \multirow{2}{*}{2,00,000} & \multirow{2}{*}{448} & \multirow{2}{*}{Google Colaboratory} & \multirow{2}{*}{96.77\%} \\
     &  &  &  &  \\ \hline
    AutoML (Static and Online) & 2,00,000 & 2381 & 2 vCPUs,448 GB RAM,16 GB of VRAM, 8 Tesla V100 GPUs. & 95.80\% \\ \hline
    Ensemble Learning (Bagging) & 1,60,000 & 2381 & \begin{tabular}[c]{@{}c@{}}Nvidia DGX Station A100 ,AMD EPYC 7742 64-core processor, \\ 4X Tesla A100 ,40GB of GPU memory, and 512 GB of DDR4 RAM.\end{tabular} & 96.56\% \\ \hline
    SVM (Linear SVC) & 1,32,000 & 2381 & - & 92.6\% \\ \hline
    \textbf{Our Approach} & \textbf{1,99,956} & \textbf{384} & \textbf{Intel Xeon 2.20 GHz CPU,30GB RAM,19.5 GB disk space} & \textbf{97.25\%} \\ \hline
    \end{tabular}
}
\label{tab:comparison_with_other}
\end{table*}

\section*{Acknowledgment}
The authors sincerely thank the Data Science and Artificial Intelligence (DSAI) Club of COEP Technological University for their invaluable support and for providing the opportunity to undertake this research. The club’s consistent guidance and its commitment to fostering innovation in data science and artificial intelligence played a pivotal role in the successful completion of this work.

\end{document}